\documentclass[a4paper]{jpconf}
\usepackage{graphicx}

\usepackage{citesort}

\begin{document}

\title{The Sun as a probe of \\ Fundamental Physics and Cosmology}

\author{I P Lopes$^1$}

\address{$^1$Centro Multidisciplinar de Astrof\'{\i}sica, Instituto Superior T\'ecnico, 
Universidade Tecnica de Lisboa , Av. Rovisco Pais, 1049-001 Lisboa, Portugal}

\ead{ilidio.lopes@tecnico.ulisboa.pt}

\begin{abstract}
The high quality data  provided by helioseismology, solar neutrino flux measurements, 
spectral determination of
solar abundances,  nuclear reactions rates coefficients among other experimental data, leads to the highly accurate prediction of the internal structure of the present Sun - the  standard solar model. In this talk, 
I have discussed how the standard solar model, the best representation of the real Sun,  
can be used to study the properties of dark matter,  for which two complementary 
approaches have been
developed: - to limit the number of
theoretical candidates proposed as the dark matter particles, this analysis complements the
experimental search of dark matter, and - 
as a template for the study of the impact of dark matter in the evolution of stars, 
which possibly occurs for stellar populations formed in regions of high density of dark matter, 
such as stars formed in the centre of galaxies and  the first generations of stars.
\end{abstract}

\section{Introduction}

In this talk, I discussed how the Sun is being used  
as a probe of fundamental physics and modern cosmology. 
As a nuclear physics community, if you remember the 30's and 40's,  
the Sun has played a key role in the progress and development of nuclear physics, namely, 
by contributing to the understanding of the basic nuclear processes that lead to the discovery  
of the nuclear chain reactions  (PP chains, CNO cycle and 3$\alpha$ reactions) 
by H. Bethe,  C. F. von Weizs\"acker and  F. Hoyle among others. Therefore, it is no surprise, 
that in particular the Sun, and  in general the other stars, can  play an important role to explore the validity 
of new fundamental laws of physics, such as testing new theories of gravity proposed as alternative to
General Gravity~\cite{2012ApJ...745...15C}, to discuss the precision of new measurements
of fundamental constants~\cite{2003MNRAS.341..721L}, to study the properties of
neutrino flavour oscillations~\cite{2013ApJ...765...14L,2013PhRvD..88d5006L},
and, as discussed in this talk, as a tool to determine 
the properties of dark matter (DM))~\cite{2013arXiv1308.4513K,2013arXiv1308.0338Z,2013arXiv1305.4939P}. 

Undoubtedly,  the origin of all matter, and which fundamental particles matter is made off,
 is one of the leading problems of modern physics. Although, our visible Universe is made off
baryons, including our own Sun and ourselves, it is by now well-established that most matter 
(and energy) of the universe is non-baryonic. Its fundamental nature is still a mystery to us. 
Presently, the unknown non-baryonic matter content of the universe is presented 
in its two components, dark matter and dark energy -- 
a split based on their two distinct gravitational effects, dark matter accounts 
for the extra gravity needed to hold baryonic matter together (example: clusters of galaxies, stars in galaxies),
and  the dark energy  must exist to explain the expansion rate of the present universe. 
Dark matter and dark energy correspond to 24\%  and 72\% of all the matter of the Universe, respectively.
It is a quite challenging task to us, trying to understand 
this new type of matter and energy based on fundamental theories of physics 
that have been developed to explain the 4\% of baryonic matter. In this short review,
I  will focus on the study of the impact of dark matter in the evolution of the Sun and stars.       
  
The presentation is organized as follows: In the next section, I make a brief summary of 
the quality of the standard solar model (SSM). In the third section, I make a synopsis 
of the current knowledge on the origin of dark matter. Finally, in the fourth and fifth sections, 
I will discuss the impact of dark matter in the Sun and stars and 
highlight a few  DM predictions that are possible to be made using stars.    

\begin{figure*}[t]
\begin{minipage}{18pc}
\includegraphics[width=18pc]{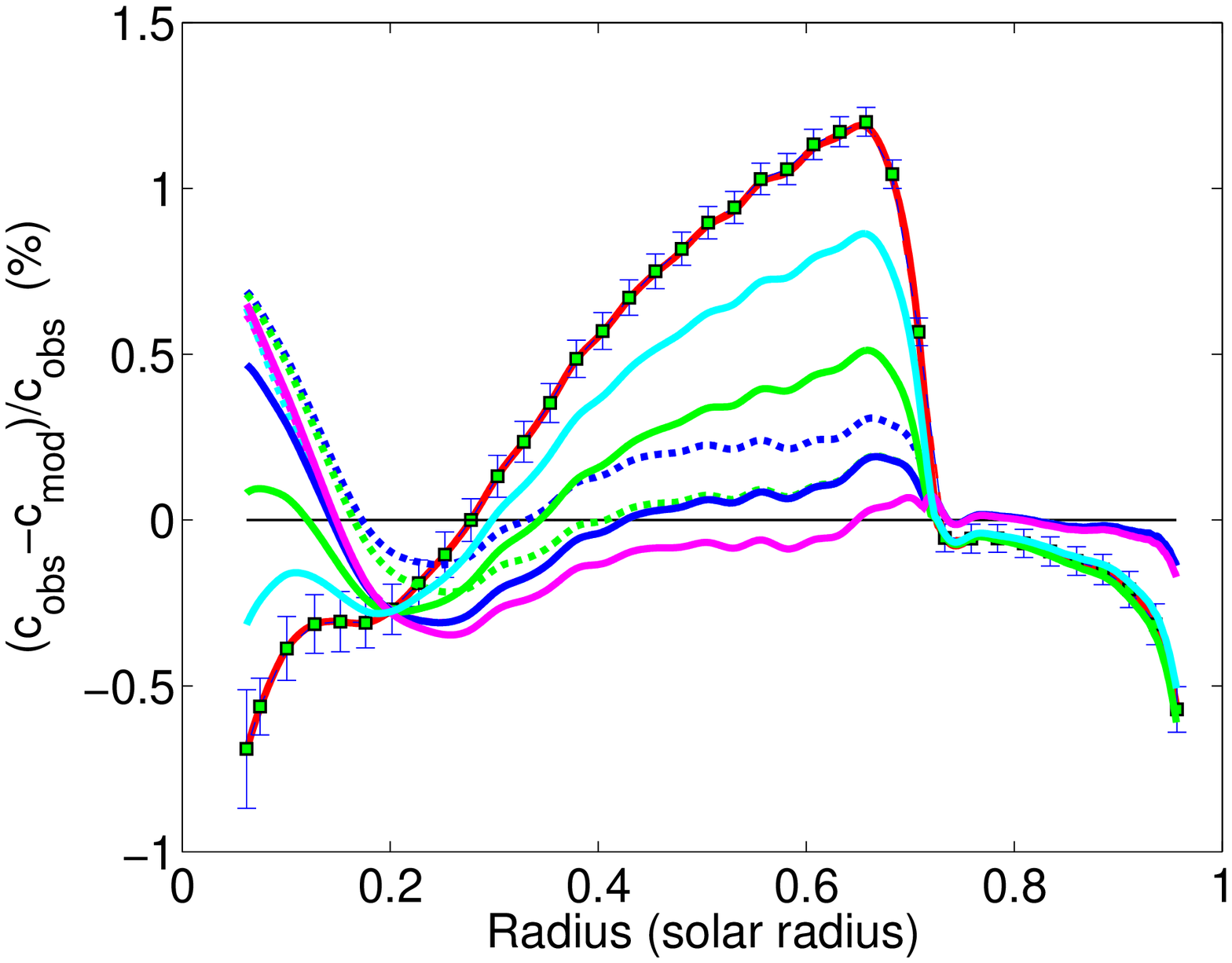}
\caption{
\label{fig:a}
\small
Relative differences between the sound speed 
inverted using the helioseismic data~\cite{1997SoPh..175..247T,2009ApJ...699.1403B},
and the sound speed deduced from the SSM (continuous red curve with error bars).
The error bars are multiplied by a factor of 10.    
This figure also shows a set of solar models with mass-loss, with 
an anomalous chemical composition. See Ref.~\cite{Lopes:2013uxa} for the details.}
\end{minipage}
\hspace{2pc}%
\begin{minipage}{18pc}
\includegraphics[width=18pc]{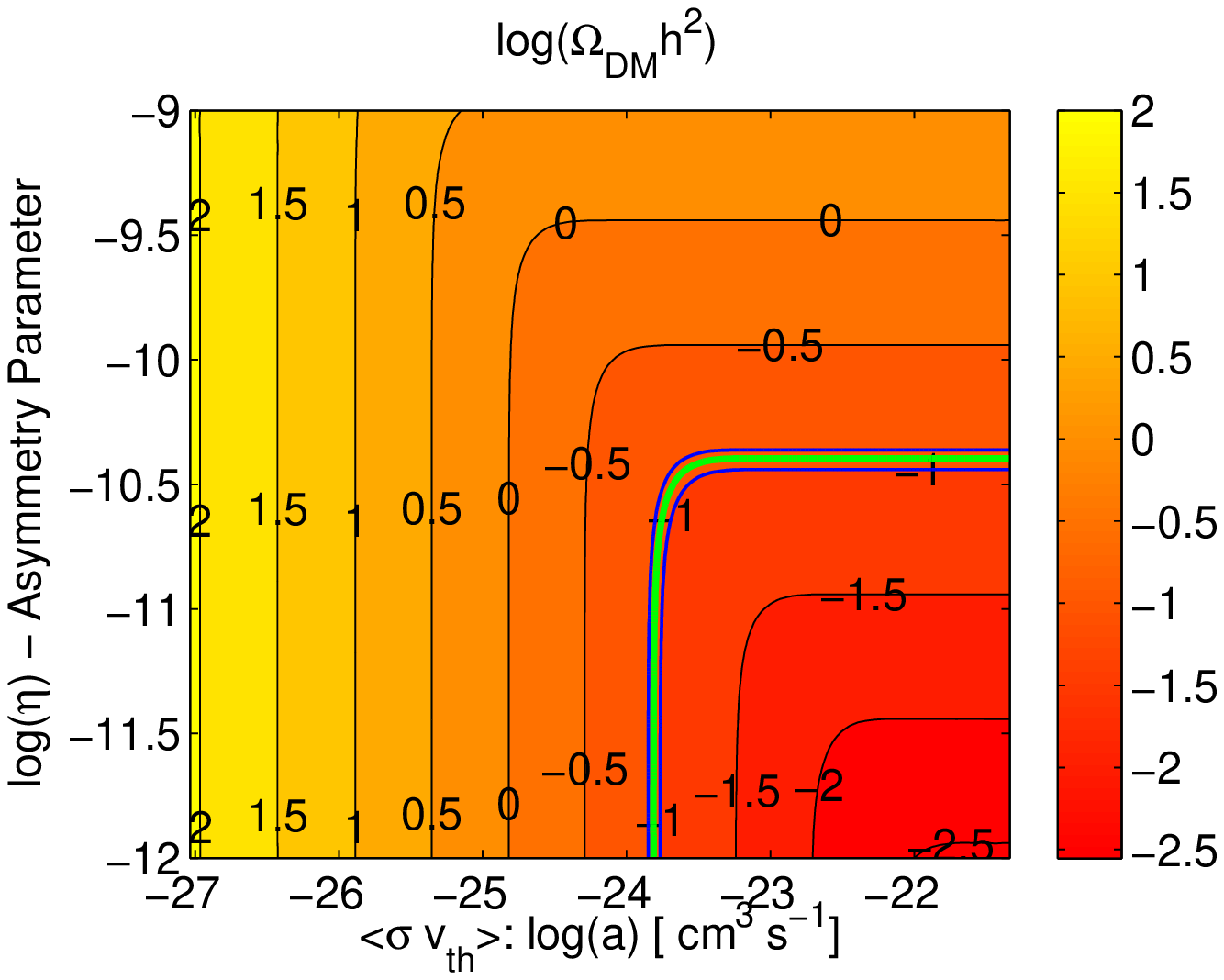}
\caption{\label{fig:b} 
\small The figure shows iso-curves of  the relic dark matter density  $\Omega_{DM}h^2$  
as a function of the $\eta_{DM}$ and $\langle \sigma v \rangle $ (s-wave annihilation channel).   
The model corresponds to a primitive universe made of DM particles with a mass of $10 \;{\rm GeV}$
($g_\chi = 2$  and $g_\star = 90$). The blue lines define the set of DM models that are
compatible with $\Omega_{DM}h^2$ observations.  
The current measurements  $\Omega_{DM}h^2$ and $\eta_{B}$ are
$0.1109\pm 0.0056$ and $(0.88\pm 0.021)\times 10^{-10}$~\cite{2011ApJS..192...18K,2012arXiv1212.5226H,2013arXiv1303.5062P}.
The figure shows that $\eta_{DM}$ and $\eta_{B}$ are of the same order of magnitude. 
See Ref.~\cite{2012ApJ...757..130L} for the details.}
\end{minipage} 
\end{figure*}

\section{The standard solar model}

The Sun is by far the best known star of the Universe. Given the advantage of being our nearest star, we have an exceptional amount of photometric and spectroscopic observational
data from the Sun. Moreover, we can probe its interior 
with  very high accuracy, by means of helioseismology and solar neutrino fluxes~\cite{1993ApJ...408..347T,2011ApJ...743...24S,Lopes:2013uxa}. 
Actually, we now know better the core of the Sun than the centre of the Earth. 
Helioseismology has obtained  a large amount of data -- more than 7 thousand acoustic modes 
were measured with a very high precision, better than one part in hundred thousand~\cite{2012RAA....12.1107T}.  During the last three decades, such high quality data has lead to a 
significant improvement of the description of plasma physics of the Sun's interior,
in particular its microscopic physics, such as the nuclear reaction network, 
the equation of state and the coefficients of radiative transfer (stellar opacities). 
In particular, Helioseismology played a leading role in the resolution of the solar neutrino problem 
(see talk of Sylvaine Turck-Chi\'eze).       

The quality of the present solar model is illustrated in Fig.~\ref{fig:a}
that shows the comparison between the radial sound speed profile obtained from  
helioseismology data and the one computed from SSM. The difference is smaller than 2\%. 
This difference although small is important. The  cause
responsible for such difference is not known yet, but very likely  
it is related with uncertainties on the coefficients of radiative transfer,
or an anomalous internal chemical composition related with the formation of the solar system~\cite{2011RPPh...74h6901T,2012RAA....12.1107T}.
Presently, the quality of the SSM  is simultaneously 
consistent with helioseismology and  solar neutrino flux data.
Although solar neutrino predictions show small differences relatively to observations, 
these are mainly related with the parameters of neutrino flavour oscillations and 
possibly small changes of the physics of the Sun's core.
 
I also have highlighted that other fields of  observational stellar astrophysics 
have contributed to this subject. 
During the recent decades, astronomers have produced high-precision catalogues 
(including high-precision spectroscopic measurements), that can be used  
to test the theory of stellar Evolution (and its new variants). 
An example is the Hipparcos Catalogue~\cite{1997A&A...323L..49P} with more than 
100,000 stars, particularly useful to study stellar populations. 
Astereoseismology, like the COROT~\cite{2007AIPC..895..201B} and Kepler~\cite{2010ApJ...713L..79K} 
missions which have observed more than one hundred pulsating's stars. 
These new sets of stellar data will significantly improve our knowledge 
about the physics of stars, but will also open the way 
to use stars as a tool for fundamental physics and cosmology studies.

In summary, although our understanding of the properties of the solar and stellar plasmas still requires some improvements~\cite{2012RAA....12.1107T}, it is clear that the Sun and sun-like stars can be used 
as tools to put constraints in fundamental physics, like dark matter particles
properties~\cite{2002PhRvL..88o1303L,2012ApJ...752..129L,2002MNRAS.331..361L,2010Sci...330..462L,2010ApJ...722L..95L,2010PhRvD..82h3509T,2010PhRvD..82j3503C,2011MNRAS.410..535C,2013ApJ...765L..21C}, as presently is done for other stars, such as neutron stars~\cite{2012PhRvL.108s1301K,2011PhRvD..83h3512K,2011PhRvL.107i1301K,2010PhRvD..82f3531K,2012arXiv1212.4075K}.
 
\begin{figure}[h]
\begin{minipage}{18pc}
\includegraphics[width=18pc]{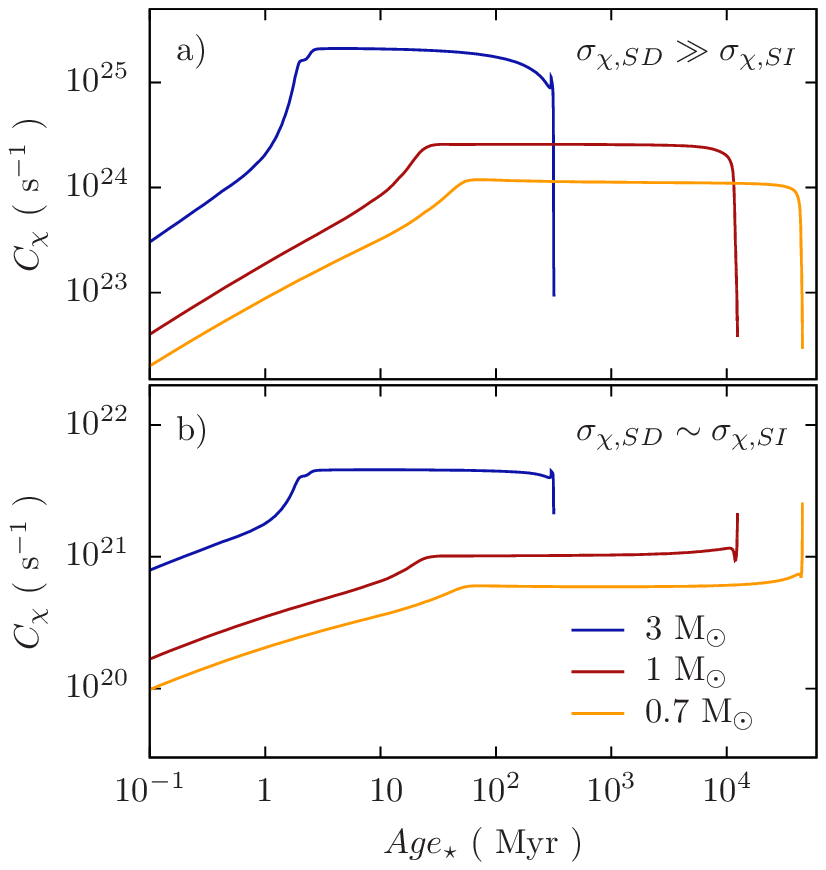}
\caption{
\small
\label{fig:c}  
Rate at which DM particles are captured during the life of stars with different masses. 
The capture rate increases during the pre-MS, is constant through the MS, and varies rapidly in the RGB. 
We assumed a halo with a density of $0.3\;$GeV$\;$cm$^{-3}$, constituted
by  DM particles with a mass of  $100\;$GeV for which  the DM-baryon scattering dominated (a) by the SD component $\sigma_{\chi,SD}=10^{-38}\;$cm$^2$ and  (b) by the SI one $\sigma_{\chi,SI}=\sigma_{\chi,SD}=10^{-44}\;$cm$^2$. 
See Ref.~\cite{2011PhRvD..83f3521L} for the details.
}
\end{minipage}\hspace{2pc}%
\begin{minipage}{18pc}
\includegraphics[width=18pc]{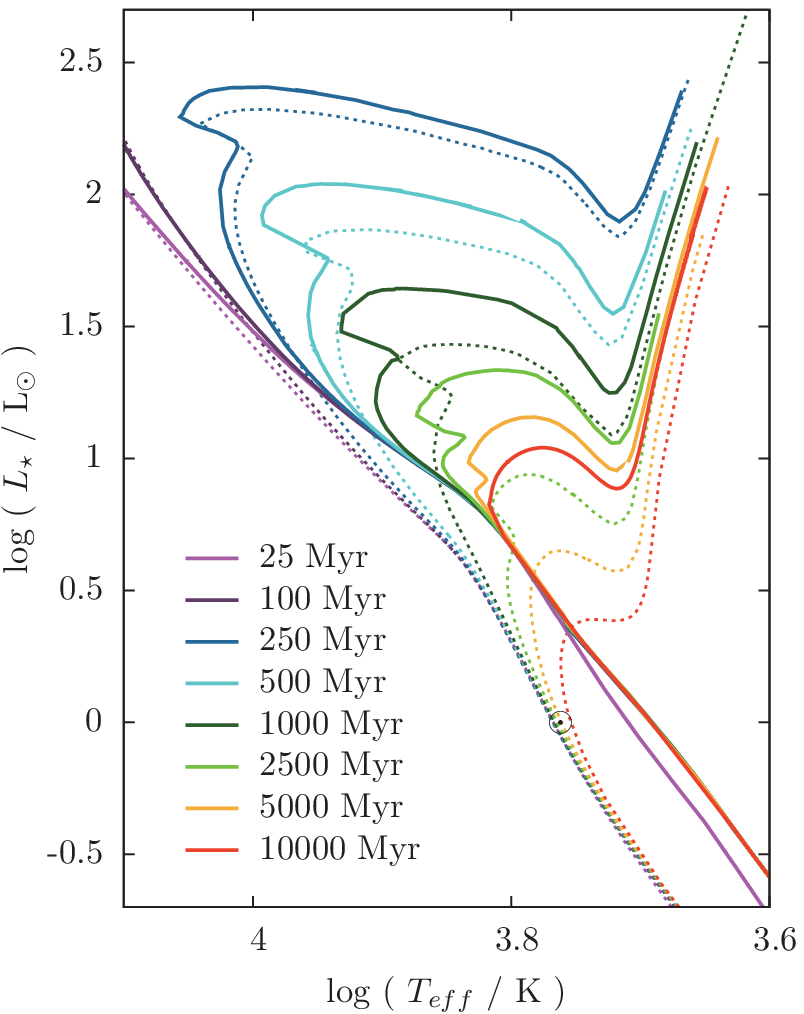}
\caption{
\small
\label{fig:d} 
Isochrones for a cluster of stars with masses between 0.7 M$_{\odot}$ and 3.5 M$_{\odot}$ 
that evolved in a halo of DM with a density $10^{10}\;$GeV cm$^{-3}$ (continuous lines) 
and for the same cluster in the classical scenario without DM (dashed lines).
See Ref.~\cite{2011ApJ...733L..51C} for the details.}
\end{minipage} 
\end{figure}

\section{The dark matter in the Universe}

Let us remind in a nutshell, what is the current status of the dark matter problem. 
Significant progress has been achieved in this subject, as a results of the strong interaction
between  Astrophysics, Cosmology and Particle Physics.

The evidence of the contribution of DM for the formation of the Universe and its structure 
is very robust. These results come from astrophysical and cosmological observations, 
as well as from numerical simulations of structure formation~\cite{2011ApJS..192...18K}. 
In many observational studies it  was found that dark matter is responsible for the 
existence of  external gravitational  field sources in different locations of the Universe,
leading to quite distinct gravitational effects, such as 
the velocity of galaxies in clusters, the rotation curves of galaxies, 
the cosmic microwave background anisotropies, the velocity dispersions of dwarf spheroidal galaxies and 
the inference of the dark matter by gravitational lensing.  
All the observational and theoretical results suggest that most of the formation of structure 
in our Universe  can only be explained by the presence of a 
gravitational field caused by the presence of a new type of particles
that must be non-baryonic and non-relativistic~\cite{2010gfe..book.....M,2013MNRAS.428.1351G}.
That means that DM particles interact with baryons through gravity and possibly
through another unknown mechanism identical to the weak interaction. DM particles 
form the gravitational web where the first stars and galaxies are created.  
In summary, these results 
strongly favour the hypothesis  that the dark matter particle is a WIMP, 
that is a Weak interacting massive particle. 

The experimental search of DM by several experimental physics collaborations 
can be organized into two groups of methods:  direct detection of DM particles --
infers the properties of the dark matter particles by the type of interaction that the
DM particles have with a target-baryon; or an indirect detection of DM particles --
by detecting one of its by-products resulting from the annihilation of DM particles, 
such as high-energy neutrinos.
A few of the experiments dedicated to dark matter searches show evidence 
of positive particle detection,  although  these results 
are still very controversial and not universally accepted:
Group  experiments such as  DAMA/LiBRA\cite{2011PhRvD..84e5014B}, 
COGENT\cite{2011PhRvL.106m1301A}, CRESST~\cite{2012PhRvD..85b1301B} 
and  CDMS II~\cite{2013arXiv1304.4279C}  
suggest that the  DM particle could have been discovered. 
These experiments show hints of a DM particle with a mass of 10 GeV  
and a cross section of $10^{-41} cm^2$.  But other experiments 
such as  XENON 10-100 \cite{2011PhRvL.107m1302A,2011PhRvL.107e1301A} exclude  
the existence of such particles.    
Nevertheless, several theoretical explanations based on the existence 
of asymmetric dark matter have been proposed to explain and accommodate all 
these positive and negative detections 
\cite{2008PhRvD..78f5040K,2010JCAP...08..018C,2011JCAP...11..010F,2011PhRvD..84h3001H,2011PhRvD..84b7301D,2011PhRvD..84k5002F,2013PhRvD..87l3507B}.
Theses hints of DM detection suggest that
at best  DM-baryon  interaction is of the order of the weak interaction.

By now, it is well-established that DM particles, like baryons, are produced 
in the primitive universe. 
Currently several theoretical DM models  suggested that DM 
is  produced by an identical process to Baryogenesis~\cite{2006PhRvD..73l3502D,2011JCAP...07..003I,2011PhRvD..84i6008D} 
-- a physical process during which an asymmetry occurs between baryons and antibaryons, 
resulting in   substantial amounts of residual baryonic matter, i.e. the
ordinary matter of the present universe. In a similar manner to Baryogenesis
for which the degree of baryonic asymmetry (unbalance between particles and antiparticles) 
is measured by $\eta_{B}$,   the asymmetry between DM particles and antiparticles
is measured by $\eta_{DM}$.  
Fig.~\ref{fig:b} shows $\Omega_{DM}h^2$ for a primitive Universe
composed by  DM particles with a mass 10 GeV~\cite{2012ApJ...752..129L}.  
As shown in this figure, only a  small set of  DM models~\cite{2006PhRvD..73l3502D,2010PhLB..687..275D}  
have a   $\Omega_{DM}h^2$ compatible with $\Omega_{DM}h^2$  
measured by WMAP and Planck experiments~\cite{2011ApJS..192...18K,2012arXiv1212.5226H,2013arXiv1303.5062P}.
Furthermore, the $\eta_{DM}$ of these DM models are of the same order  of magnitude of the observed $\eta_{B}$. Actually, there is no special reason for such coincidence. This suggests that DM and baryonic matter are more closely  connected than expected.  
In the same figure, it is worth noticing that symmetric DM  corresponds to low values  of $\eta_{DM}$  
and asymmetric DM corresponds to large values of $\langle \sigma v \rangle $.

\section{The dark matter inside the Sun and stars}
 
The numerical computation of the impact of dark matter in the evolution of the Sun 
and stars has progressed considerably in the last two 
decades~\cite{2010PhRvD..82j3503C,2010PhRvD..82h3509T,2012ApJ...752..129L},
however,  there are still a few caveats that must be addressed~\cite{2012RAA....12.1107T}. 
       
When a DM particle crosses a star like the Sun, it very occasionally scatters off a proton,
or an heavier nucleus including helium, carbon, oxygen and iron.
The scattering interaction with hydrogen is predominately spin-dependent ($SD$) and 
for the other elements spin-independent  ($SI$), 
accordingly to the scattering cross-sections $\sigma_{\chi,SD}$  and $\sigma_{\chi,SI}$, respectively. 
The interaction of DM with baryons 
depends strongly of the mass  and velocity of the DM particle. 
In certain cases, after collision  the  DM particle will lose enough energy 
to become gravitationally bound to the star, and by a sequence of orbital paths 
around  the centre of the star, the DM particle sinks to its
core~\cite{2002MNRAS.331..361L,2002MNRAS.337.1179L,2011PhRvD..83f3521L}. 
The total amount of particles  (and antiparticles) of DM inside of the star 
at a given time is regulated by three fundamental processes:  
capture of DM particles from the DM halo, the 
evaporation of DM particles from the star (in the Sun only relevant for DM particles 
with a mass smaller  than 5 GeV)  and the annihilation of DM particles.
It is the balance between these three processes that determines the total amount of DM
inside the star and consequently the influence  that DM has on the 
evolution of the host star.  
Fig.~\ref{fig:c} show the variation of the dark matter capture rate for different stars 
in different phases of their evolution. 
The capture rate of dark matter depends largely of the radius 
of the star at each time step evolution. The chemical composition of the star 
also plays a major role in the capture of DM by the star, 
for which its content in metals (chemical elements other than hydrogen and helium) 
determines which is the dominant type of scattering~\cite{2011PhRvD..83f3521L}.

Once captured,  DM particles can change the evolution of the star through two mechanisms:   
(i) by supplying the star with a new source of energy due to DM annihilation --
which complements the PP chain, CNO cycle and 3$\alpha$ nuclear reactions
in balancing the self-gravitational contraction of the star,  or (ii) 
by providing  a new mechanism for the transport of energy -- this  complements
the radiative and convective energy transport.
The two  processes can work simultaneously, however, in most of the cases when  
the evolution of the stars is affected by the presence of a DM energy source, 
the DM energy transport can be neglected. 
The first mechanism is important in locations of the Universe with
high DM density, such as the regions where the first stars and galaxies are formed,
and  the center of galaxies like our Milky Way and the  spheroidal galaxies.  
The second mechanism becomes relevant in much less denser DM regions  
like in the Sun's neighbourhood.

The density of the DM halo where the star is formed is one of the major 
sources of uncertainty. 
Usually the DM density is assumed to be  $0.38 \; {\rm GeV cm^{-3}} $
~\cite{2010JCAP...08..004C}.
Nevertheless,  recently estimations found a DM density  of the order of    
$ 0.3 \; {\rm GeV cm^{-3}} $~\cite{2012ApJ...756...89B}
or  $ 0.85 \; {\rm GeV cm^{-3}} $~\cite{2012MNRAS.tmp.3493G}. 
This discrepancy has an important impact on our study, nevertheless,
researchers in the field use the lower DM density value 
to make more conservative predictions of DM parameters.  
However, there are other sources of uncertainty in these computations, 
which we must resolve  to obtain more reliable predictions. 
Among others, here is a list of a few that I believe   must be 
investigated:  
(i)
Particle Physics and Nuclear Physics: properties of the DM particles,
like   the mass of DM particle, the scattering cross-section of the DM with baryons, 
annihilation rate of DM particles and antiparticles, annihilation channel
and nuclear factors.    
(ii)
Cosmology: The dynamics and evolution of the dark matter halo, 
in particular the density of the DM halo where the star is formed, 
the dynamical properties of the DM particles like their thermal velocity.  
(iii)
Astrophysics and Stellar Physics: The velocity of the star, the detailed internal structure of the star, 
like its chemical composition. Although, some of these properties
are well known in the case of the Sun, it is not yet the case for other stars. 

Although, we found in preliminary tests~\cite{2011PhRvD..83f3521L} that such caveats have a relative small effect in the accretion of DM by stars,    as this research field progresses, a better account of such mechanisms must be included in the  models. 

\begin{figure}[h]
\begin{minipage}{18pc}
\includegraphics[width=22pc]{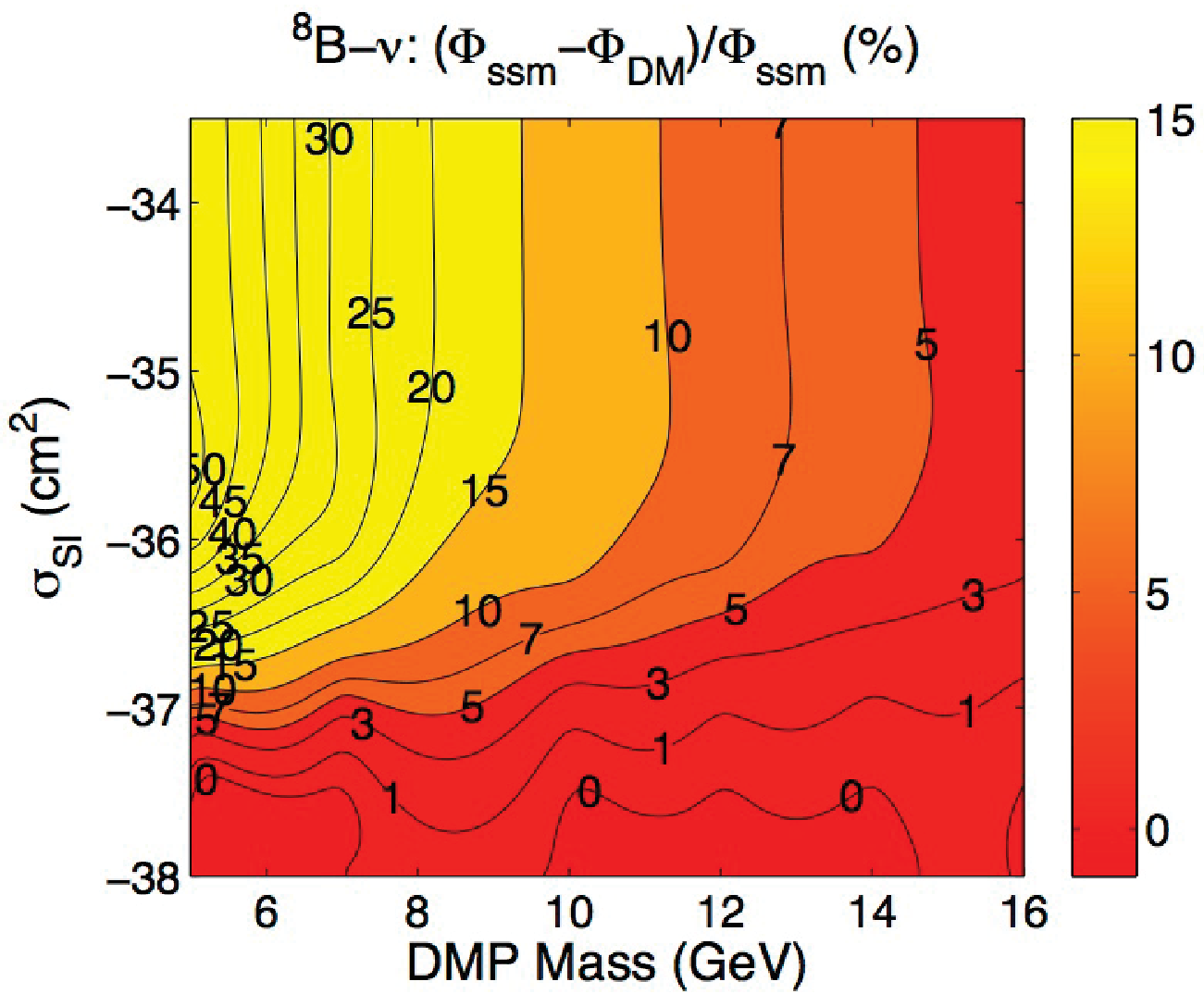}
\caption{
\small
\label{fig:e}  
Percentage decrease changes of the solar $^8$B neutrino flux $\Phi_\nu(^8B)$
for the Sun evolving within different DM halos relatively to the SSM: 
The properties of the
DM particles and the stellar physics can be found in the Ref.~\cite{2012ApJ...752..129L}. 
}
\end{minipage} 
\hspace{2pc}%
\begin{minipage}{18pc}
\includegraphics[width=18pc]{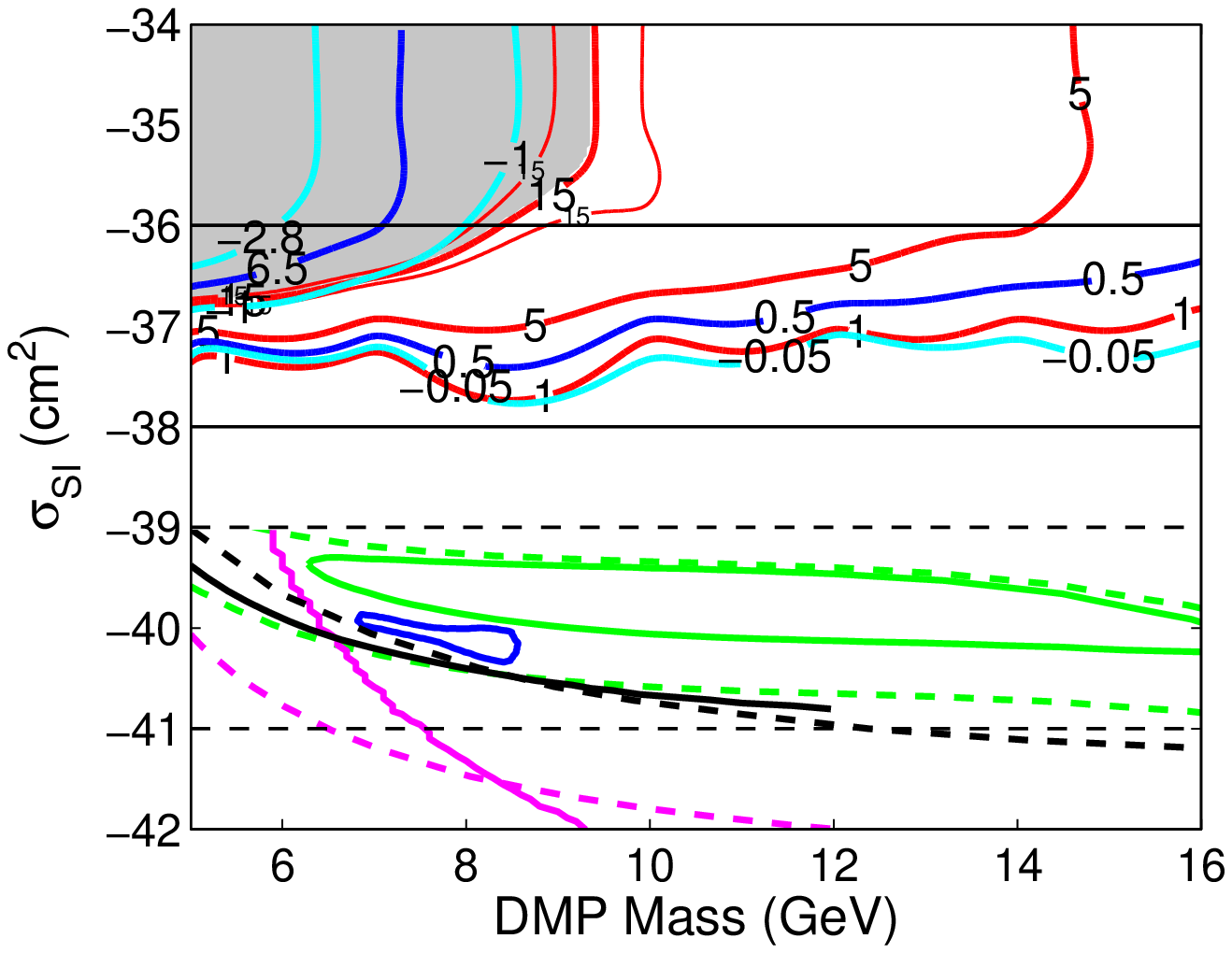}
\caption{ 
\scriptsize
\label{fig:f} 
Exclusion Plot of DM searches: 
{\bf (1)} CoGeNT (blue curve), DAMA/LIBRA a $3\sigma$ CL  (solid green curve) and $5\sigma$ CL 
(dashed green curve); XENON 10 (dashed magenta curve), XENON100 (solid magenta curve); 
CDMS II limit (solid black curve) and SIMPLE limit (dashed black curve).
{\bf (2)} Solar neutrino flux variations due to DM: $\Phi_\nu(^8B)$ (solid red curve),   
$\Phi_\nu(^7Be)$ (solid blue curve)  and $\Phi_\nu(pep)$ (solid cyan curve).
The grey region corresponds to values of  $\Phi_\nu(^8B)$ neutrino flux larger than  $15\%$ (red curve). See Ref.~\cite{2012ApJ...752..129L} for the details. 
}
\end{minipage}
\end{figure}

\section{Dark matter constraints from Sun and stars}

In the following, I provide a few examples for which DM has quite distinct and important 
effects in the evolution of stars. 

{\it The case of a star evolving in a DM of high density}: 
the gravitational contraction of a star during the pre-main sequence phase evolving within 
DM halo is identical, but not equal to a normal star.  In the case of a star forming  
in a normal molecular cloud the self-gravity of the star leads 
to a rapid contraction with an increase of its internal temperature, the star
follows a well defined path in the H-R diagram~\footnote{Hertzsprung-Russell (H-R) Diagram: Luminosity vs.
effective temperature of stars -– each star of a fix mass describes a unique path in this diagram.}.  
The evolution of this star within a dense DM halo is almost the same, nevertheless due to the extra 
source of energy provided by DM annihilation,
the star gravitational contraction progresses at a much lower time step. 
This DM effect has a major implication  for stellar clusters, 
namely by changing the location of main-sequence of stars, 
and the turnoff point~\footnote{
The turnoff point for a star refers to the point on the H-R diagram, where the star leaves the main-sequence after the exhaustion of its main fuel.} at which stars leave the main-sequence. This effect is shown in  
the isochrones\footnote{Luminosity-temperature plot of a cluster of stars with different masses at a fixed age.} of a stellar population formed in a high density DM halo. In particular, I notice that   
{\it the main sequence of low mass stars} 
has shifted to a high value of Luminosity when compared with a normal stellar cluster (cf. Fig.~\ref{fig:d}).  
If such type of isochrones are discovered in a stellar cluster on first generation of stars
or at the core of galaxies including in the Milky-Way, then this is a 
powerful indication that dark matter is contributing for their evolution~\cite{2009ApJ...705..135C,2011ApJ...742..129S,2011ApJ...729...51S,2012MNRAS.422.2164I}.
    
{\it The case of a star evolving in a DM of low density}:   
Another type of impact of dark matter in stars, like the Sun, occurs for stars formed in the disk of galaxies,
where the DM density is much smaller than in the case previously discussed. 
For a star like the Sun, the capture of DM  depends strongly of the scattering 
cross section with baryons: If the DM particle has a relatively large scattering cross section with Hydrogen 
(spin dependent cross section) or with heavy elements (like Oxygen, Iron among others, spin independent),
the number of collisions between DM particles and Baryons is high, consequently in the extreme cases 
this leads to the formation of an isothermal core~\cite{2002PhRvL..88o1303L}, otherwise this effect
only reduces the central temperature of the star. By using the current probes of the solar
interior, such as solar neutrinos, we are able to put constraints on DM parameters.
At present, only 3 sources of solar neutrinos (like $^8$B, $^7$Be ad PeP) 
are observed by the current solar neutrino detectors  (Borexino, SNO and Kamiokande), 
for which all are sensitive to the temperature of the Sun's core. 
Fig.~\ref{fig:e} shows the DM impact on $^8$B neutrino flux for DM solar models, 
for which the $^8$B neutrino flux is compared with the 
$^8$B flux of the SSM. This example shows DM particles
for which the $^8$B neutrino flux variation is larger than 15\%
which is larger than the current uncertainty of the SSM, 
therefore such DM particles can be excluded~\cite{2010Sci...330..462L}.  
These DM constraints are based on the hypothesis that the $^8$B neutrino flux difference 
between the observed and predicted SSM  is smaller than 15\%. 

Fig.~\ref{fig:f} shows a second interesting example, which uses data of solar neutrino flux measurements:
Feng {\it et al.}~\cite{2011PhLB..703..124F} have shown the previous experimental data of DM search (including the exclusion limits – considering the usual interaction between DM and the target baryons),
could be reconciled if one considers that  interaction of the  DM particle with the  target baryon 
occurs assuming  iso-spin violation.  As a consequence the spin-independent scattering cross-section 
of the  DM-target Baryon increases from $10^{-40} cm^2$ to $10^{-37} cm^2$. Lopes and  Silk~\cite{2012ApJ...752..129L} have shown that  $^8$B neutrino fluxes can reject the lower masses 
and higher values of scattering cross section. Further future $^8$B neutrino flux measurements
could lead to much better constraints  (cf. Fig.~\ref{fig:f}).

I would like to notice that this type of analysis could be strongly improved if we could use 
other sun-like stars of different masses from the Sun for which the impact of dark matter could be more visible. In a recent article Casanellas and Lopes~\cite{2013ApJ...765L..21C} use oscillation spectra of the stars 
HD 52266 (1.1 $M_\odot $)  and alpha Centari B (0.9 $M_\odot $) to put constraints in DM
(see poster Casanellas and Lopes).  In particular, they found that particles with a mass of $5 GeV$
and a spin-dependent scattering cross-section  smaller than $3\times 10^{-36} cm^2$  
can be excluded with a 5-$\sigma$ level. 

Finally, I hope that in this talk I have convinced you that the solar and stellar physics 
and their related fields  of  nuclear physics can provide fundamental tools to test the  
different experimental and  theoretical DM particle candidates. 
The author thanks the Astrophysics Portuguese Society (SPA) and  the organisation for the invitation to participate in the conference. 
 
\section*{References}
\providecommand{\newblock}{}

\end{document}